\documentclass[conference]{IEEEtran}

\usepackage{enumerate}
\usepackage{multirow} 
\usepackage{graphicx}
\usepackage{subcaption}
\usepackage{pgfplots}
\usepackage{tikz}
\usetikzlibrary{shapes,arrows,positioning}
\usepackage{color}
\usepackage{todonotes}
\usepackage{colortbl}
\usepackage{changes}
\usepackage{flushend}
\usepackage[T1]{fontenc}
\usepackage[nolist]{acronym} 
\usepackage[miktex]{gnuplottex}
\usepgfplotslibrary{external} 
\usepackage{url}
\usepackage[pdfusetitle, pdfauthor={Marcel Kneib (Robert Bosch GmbH), Oleg Schell (Bosch Engineering GmbH), Christopher Huth (Robert Bosch GmbH)}]{hyperref}

\begin{document}

\title{On the Robustness of Signal Char\-ac\-ter\-is\-tic-Based Sender Identification}

\author{\IEEEauthorblockN{Marcel Kneib}
\IEEEauthorblockA{Robert Bosch GmbH\\
Mittlerer Pfad 9\\
70499 Stuttgart, Germany\\
Marcel.Kneib@de.bosch.com}
\and
\IEEEauthorblockN{Oleg Schell}
\IEEEauthorblockA{Bosch Engineering GmbH\\
Robert-Bosch-Allee 1\\
74232 Abstatt, Germany\\
Oleg.Schell@de.bosch.com}
\and
\IEEEauthorblockN{Christopher Huth}
\IEEEauthorblockA{Robert Bosch GmbH\\
Robert-Bosch-Campus 1\\
71272 Renningen, Germany\\
Christopher.Huth@de.bosch.com}
}

\maketitle
\begin{acronym}
	
	\acro{ADC}{analog-digital-converter}
	\acro{CAN}{Controller Area Network}
	\acro{CAN-FD}{CAN with flexible data rate}
	\acro{ECU}{Electronic Control Unit}
	\acro{FN}{False Negative}
	\acro{FP}{False Positive}
	\acro{IDS}{Intrusion Detection Systems}
	\acro{IG}{Information Gain}
	\acro{LR}{Logistic Regression}
	\acro{NB}{Naive Bayes}
	\acro{MAC}{Message Authentication Code}
	\acro{OBD}{on-board diagnostics}
	\acro{SecOC}{Secure Onboard Communication}
	\acro{SVM}{Support Vector Machine}
	\acro{Viden}{Voltage-based attacker identification}
	\acro{HSM}{Hardware Security Module}
	\acro{MCU}{microcontroller}
	\acro{FPU}{floating point unit}
	\acro{GTM}{Generic Timer Module}
	\acro{DSO}{Digital Storage Oscilloscope}
	
\end{acronym}

\begin{abstract}
	Vehicles become more vulnerable to remote attackers in modern days due to their increasing connectivity and range of functionality.
	Such increased attack vectors enable adversaries to access a vehicle \ac{ECU}. 
	As of today in-vehicle access can cause drastic consequences, because the most commonly used in-vehicle bus technology, the \ac{CAN}, lacks sender identification.
	With low limits on bandwidth and payload, as well as resource constrains on hardware, usage of cryptographic measures is limited. 
	As an alternative, sender identification methods were presented, identifying the sending \ac{ECU} on the basis of its analog message signal. 
	While prior works showed promising results on the security and feasibility for those approaches, the potential changes in signals over a vehicle's lifetime have only been partly addressed.   
	This paper closes this gap. 
	We conduct a four months measurement campaign containing more than 80,000 frames from a real vehicle. 
	The data reflects different driving situations, different seasons and weather conditions, a 19-week break, and a car repair altering the physical \ac{CAN} properties. 
	We demonstrate the impact of temperature dependencies, analyze the signal changes and define strategies for their handling.
	In the evaluation, the identification rate can be increased from 91.23\,\% to 99.98\,\% by a targeted updating of the system parameters.
	At the same time, the detection of intrusions can be improved from 76.83\,\% to 99.74\,\%, while no false positives occured during evaluation.
	Lastly, we show how to increase the overall performance of such systems by double monitoring the bus at different positions.
\end{abstract}

\section{Introduction}
\label{sec_intro}
Modern vehicles have an increasing amount of functions and services with an increasing amount of connectivity to their environment~\cite{Koopman2006, Hoppe2008, Ring2017, Szilagyi2019}.
Local wireless communication uses standards like Bluetooth or WiFi, but also cellular radio is utilized to communicate with cloud systems.
The additional functionalities and services also open up new attack vectors, all without requiring physical access at any time to vehicles~\cite{Checkoway2011, Studnia2013}.
By demonstrating vulnerabilities on such vehicles, numerous publications show that this threat is real~\cite{Higbee2007, Koscher2010, Miller2013}.
The most famous example in this respect is the attack on the Jeep Cherokee~\cite{Miller2015}.
Using a wireless connection, the researchers were able to manipulate safety-critical functions such as braking and steering.
That a vulnerability in an \ac{ECU} can be exploited in such a way is primarily due to the lack of security mechanisms~\cite{Hoppe2008} in the most commonly used standard for internal vehicle communication, the \ac{CAN}~\cite{CAN1991}.
Once an attacker has acquired access to the vehicle communication, he can potentially control numerous vehicular functions.
Even though the attack on the Jeep dates back several years, the current work of Tencent Keen Security Lab shows that these dangers still exist in more modern vehicles~\cite{Keen2018, Keen2019}.
The team discovered several vulnerabilities in different BMW models that allow access to the \ac{CAN} via wireless connection. 

The general use of cryptographic measures to secure the in-vehicle communication is limited due to the constrained resources of the \acp{ECU}, the low bandwidth and payload per \ac{CAN} message~\cite{Wolf2007, Lin2012, Groza2013, Kneib2018}.
Therefore, alternative approaches have been presented in the past which allow to identify the sender of a \ac{CAN} message based on physical characteristics of its analog signal \cite{Murvay2014, Cho2016, Choi2016, Choi2018, Kneib2018}.
In this regard, the general idea is to extract a fingerprint from a measured signal, which can then be compared with known patterns to determine the sending \ac{ECU}.
Since the in-vehicle communication is of static nature, authorization of an identified \ac{ECU} to send a message can be validated. 
Besides detection of forged frames, these techniques can also enrich comprehensive \ac{IDS} with relevant sender information \cite{Hoppe2008, Mueter2010}. 
Such systems can be roughly separated into two groups: signature-based systems which detect intrusions based on predefined rules and anomaly-based systems which are capable of detecting new attacks based on irregularities in the communication.

\subsection{Contribution}
Regardless of the approach of signal characteristic-based sender identification, the related work on this topic pertain roughly the same concept. 
However, there are open questions regarding stability during operation \cite{Murvay2014, Cho2016, Choi2016} which are answered in a limited extent~\cite{Choi2018, Kneib2018}.
This work closes this gap. 
We clarify the issues of signal characteristic stability in the course of system operation and present more precise possibilities to deal with potential changes during runtime.
For a general evaluation of the long-term deviations of the signals, a method for calculating the signal deviation is introduced and used for the analysis.
This allows a representation of the temperature dependence as well as a general behavior of the signals during operation of the vehicle.
Furthermore, different learning strategies for the creation of the initial model and a mechanism for the recognition of changes and the adaptation of the model to them were evaluated. 
While in Scission\,\cite{Kneib2018} the update mechanism is only explained very briefly, with this work we specify its implementation and evaluate its performance with regard to different signal variations. 
In addition, this work shows that the robustness of the fingerprinting techniques can be increased by using several measurement points.
Therefore, the approach for sender identification and intrusion detection, Scission~\cite{Kneib2018}, is used during the evaluation of this work.
For the experiments, several data sets were gathered, which were recorded at different seasons and temperatures as well as at different conditions of the vehicle.
In total, the data contains more than 80,000 frames, collected over a period of more than 4~months.
The data also includes a 19-week break during which signal-changing repairs were made to the vehicle.
Through the methods presented in this paper we achieve an identification rate of over 99.9\,\%, an attack detection rate of 99.7\,\% as well as zero false positives and thus keep the performance in comparison to Scission~\cite{Kneib2018} also with changes of the signals.

\subsection{Organization of the Paper}
After the introduction in Section\,\ref{sec_intro}, Section\,\ref{sec_background} provides background information on the \ac{CAN}, as well as reasons for the variation of the signal characteristics.
In addition, this section provides an overview of the sender identification and intrusion detection approach used in this paper, as well as an explanation of the significance of changes of input parameters.
In Section\,\ref{sec_signal_char_analysis}, the actual signal changes are examined independently of the employed identification procedure.
This includes the calculation as well as the analysis of signal deviations based on data of a real vehicle.
Subsequently, the evaluation in Section\,\ref{sec_evaluation} follows with regard to robustness and performance of different strategies for handling signal changes in combination with sender identification. 
After having presented the results of the investigations, Section\,\ref{sec_related_work} addresses the analysis and evaluation of the existing sender identification mechanisms with regard to the handling of changes.
We conclude the paper in Section\,\ref{sec_conclusion}.

\begin{figure*}[t!]
	\centering
	\includegraphics[width=\textwidth]{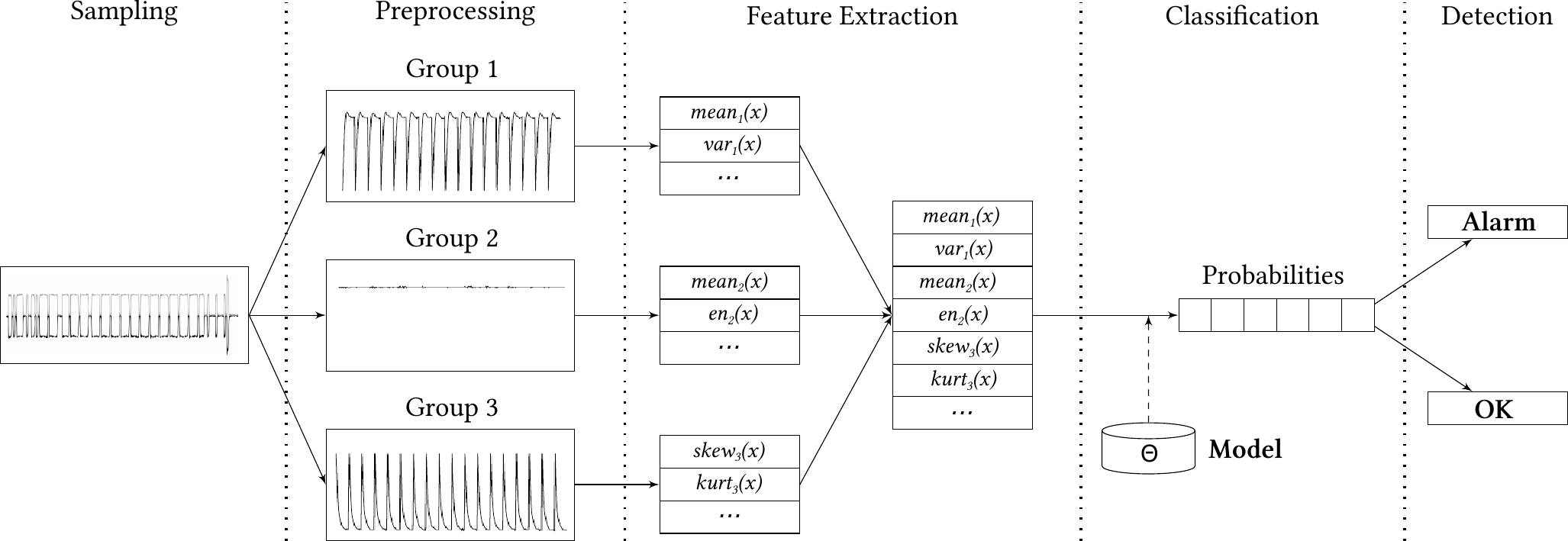}
	\caption{General procedure of the intrusion detection system Scission \cite{Kneib2018}.}
	\label{fig:scission_approach}
\end{figure*}

\section{Background}
\label{sec_background}
\subsection{Controller Area Network}
Representing an often used bus system, the CAN\,\cite{CAN1991} enables data exchange of vehicular ECUs in a broadcasting communication.
It was designed that transmitted messages contain an unique identifier to characterize each message.
Further, CAN is built as a multi-master network, causing that several control units may try to initiate a data transmission at the same time.
In this case, the arbitration process determines which ECU is entitled to occupy the bus based on the priority of the message to be sent, which in turn is defined by its identifier.
On a physical level, the transmission takes place over two twisted wires, CAN high and low, which are connected to the transceiver of each ECU and are terminated at the ends with 120$\Omega$.
Logical bits are transmitted over those two wires by their differential voltage.
In simplified terms, this procedure is described in the following.
During the transmission of a logical one, both channels rest at a voltage of 2.5\,V resulting in a difference of 0\,V.
Should a zero bit be transmitted, CAN high is driven towards 3.5\,V whereas CAN low is pulled to 1.5\,V, leading to the fact that a logical zero is represented by 2\,V in the differential signal.
However, the theoretical target voltage of 2\,V and the signal shape during bit transitions slightly differ between the ECUs, even though they share the same bus and are powered by the same 12\,V/24\,V battery.
In order to derive the ECU operation voltage of 5\,V and to prevent fluctuations of the same, voltage regulators are brought into operation.
Due to manufacturer-related imperfections of electronic components built in both the regulators and the transceivers, small variations of the voltage during transmission are likely to happen for each ECU~\cite{Mittal2016}.   
The same applies to the grounding voltage~\cite{Vector2003}, which may vary due to external influences like temperature or humidity.
Additionally, the resulting voltage on the wire is affected by reflections of the power of the transmitted signal~\cite{Mori2012}.
These occur due to cable characteristics like channel length and bus termination.
To summarize, all the aforementioned factors have an impact on both the resulting voltage level during the transmission of a zero bit and the shape of the transitions between bits of opposed values.

\subsection{Scission}
Existing differences of voltage characteristics can be exploited to realize an authentication mechanism for the connected ECUs.
An approach, which realizes this idea to provide security for the CAN is called Scission\,\cite{Kneib2018}.
By utilizing machine learning techniques to differentiate the voltage variations of each ECU, the system can draw conclusions on the origin of a transmitted CAN message.
For this purpose, Scission initiates its authentication procedure with the sampling of voltage values during the transmission of a CAN message as visualized in Figure\,\ref{fig:scission_approach}.
Afterwards, the sampled signal is separated into three different groups, from which the first group contains all rising edges which occurred during the transition from a logical one to a zero within the data field of the respective message.
Further, the second group holds the voltage values during the transmission of a logical zero, whereas the last group, analogously to the first, stores all falling edges.
To improve the performance of machine learning approaches in general, it is advised to transform the data into meaningful representations, called features.
Regarding Scission, a variety of different features, both in the time and frequency domain, are determined based on the voltage samples contained in the three aforementioned groups.  
With the help of this calculated feature set and an appropriate machine learning approach, a model of the mapping between available control units and their individual voltage characteristics can be established.
Since in this context, the Logistic Regression is a simple yet powerful machine learning approach and also provides great capabilities to adapt an already established model, it was chosen for the implementation in Scission.
This adaption capability is an especially important basis for the update procedure described in Section \ref{sec:update_mechanism}.
To realize the previously mentioned sender identification, the model, which in the case of the Logistic Regression consists of as many classifiers as connected ECUs, is used during the classification to determine the probability of each ECU being the origin of the message.
Here, the ECU whose classifier yielded the highest probability is chosen to be the initiator of the transmission.
In order to increase the system robustness and simultaneously decrease false alarms in case of minor uncertainties, Scission implements two thresholds to make a final decision on intrusions.
In this context, a transmission is classified as authorized, when the classifier of the eligible control unit crosses a lower probability threshold which is defined in advance.
Only if the lower threshold is undercut by the probability of the theoretically authorized ECU and an upper threshold exceeded for any other probability, an adversarial transmission originating from an unauthorized control unit is declared.
In every other case, the message is marked as suspicious.

\subsection{Concept Drift}
\begin{figure}[h!]
	\centering
	\includegraphics[width=\linewidth]{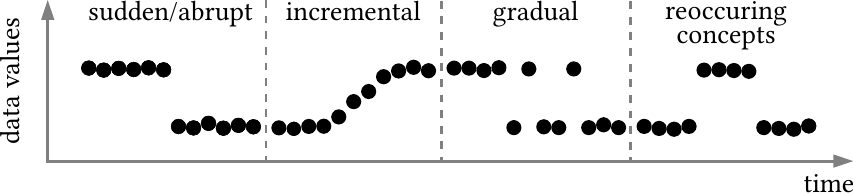}
	\caption{Different kinds of concept drift \cite{Gama2014}.}
	\label{fig:concept_drift}
\end{figure}
Machine learning approaches are capable of determining the relationship between structures in the provided training data and their target values.
In this process, it is generally assumed that the characteristics of the provided data during the model establishment phase do not change over the period of system utilization. 
However, in a realistic environment this assumption cannot be met as external influences dynamically change the distribution of the underlying data.
This shift in either the data characteristics or the target value is also referred to as concept drift.
As a direct consequence of this drift, the classification performance of the system deteriorates on account of an outdated model.
Here, the extent of performance loss further depends on the severity and type of the respective drift.
A differentiation can be made between an abrupt change of the distribution or an incremental drift which can be recognized and on which the system can react to.
The authors of \cite{Gama2014} comprehensively address the topic of concept drifts, whereas Figure\,\ref{fig:concept_drift} briefly illustrates both abrupt and incremental changes.
In the automotive area, material wear and temperature fluctuations may be the cause for the later drift, whereas modifications on either the communication network or the power supply could depict a reason for abrupt voltage changes.
Therefore, to maintain a good classification performance, the model either has to be adapted accordingly or it has to be rebuilt from scratch with present data.

\section{Signal Characteristic Analysis}
\label{sec_signal_char_analysis}
This section addresses the analysis of changes in voltage characteristics which were observed over a longer period of time.
For this purpose, the recorded data sets utilized for this analysis are presented first.
Based on this data and the contained CAN messages, the analytical approach is described subsequently with which the individual data sets and changes are measured.
In this respect, the results of the analysis are also presented.
Finally, the update procedure is introduced which is used to adapt and update the machine learning model for the succeeding evaluation.

\begin{center}
	\begin{table}
		\caption{Data sets for the evaluation on signal robustness.}
		\begin{tabular}{>{\centering}p{0.8cm} >{\centering}p{0.8cm} >{\centering}p{0.9cm} >{\centering}p{0.9cm} p{3.3cm}}
			\hline \\[-1em]
			\multicolumn{5}{c}{Data Set 1 - Summer}  \\ \\[-1em]
			Part & Frames & Temp. & Break & \multicolumn{1}{c}{Properties} \\
			\hline \\[-1em]
			Part 1 & 5977 & 25$^\circ$C (77$^\circ$F) & 3 min & Switched off and connected to external battery \\
			Part 2 & 4645 & 25$^\circ$C (77$^\circ$F) & 15 min & Switched off \\
			Part 3 & 6823 & 32$^\circ$C (89.6$^\circ$F) & 225 min & Switched on and following cooling down at 23$^\circ$C (73.4$^\circ$F) \\
			Part 4 & 4950 & 36$^\circ$C (96.8$^\circ$F) & 132 days & Switched off and on with including motor start\\
			& & & \\
			\hline \\[-1em]
			\multicolumn{5}{c}{Data Set 2 - Winter} \\ \\[-1em]
			Part & Frames & Temp. & Break & \multicolumn{1}{c}{Electrical Consumers} \\
			\hline \\[-1em]
			Part 1 & 5424 & 10$^\circ$C (50$^\circ$F) & 20 hrs & Wipers, turn signals and headlights \\
			Part 2 & 6794 & $\sim$5$^\circ$C (41$^\circ$F) & 106 min & Window lifters, turn signals, headlights and start-stop system\\
			Part 3 & 5503 & " & 95 min & \multicolumn{1}{c}{"} \\
			Part 4 & 6519 & " & 22 hrs & \multicolumn{1}{c}{"} \\
			Part 5 & 5937 & 5$^\circ$C (41$^\circ$F) & 91 min & \multicolumn{1}{c}{"} \\
			Part 6 & 6748 & " & 18 hrs & \multicolumn{1}{c}{"} \\
			Part 7 & 6413 & 2$^\circ$C (35.6$^\circ$F) & 245 min & Heating and start-stop system\\
			Part 8 & 6324 & " & 20 hrs & \multicolumn{1}{c}{"} \\
			Part 9 & 6590 & 0$^\circ$C (32$^\circ$F) & - & Start-stop system \\ 
			& & & \\
		\end{tabular}
		\label{tab:data_sets}
	\end{table}
\end{center}

\subsection{Data Sets}
\label{sec:data_set}
The data used for the analysis is taken from a Fiat\,500, the vehicle which was also used in the evaluation of Scission~\cite{Kneib2018}.
In this vehicle, a total of six \acp{ECU} communicate over the considered \ac{CAN} bus, whereby transmitted messages were recorded via a PicoScope\,4227 \ac{DSO} using the \ac{OBD}-II port in the front of the car.
Overall, two data sets were recorded in this setup, one of them during the summer and the other in winter.
Thus a total temperature range between 0$^\circ$C (32$^\circ$F) and 36$^\circ$C (96.8$^\circ$F) is considered.

The first data set consists of four parts, whereby three of them were already used for the evaluation of Scission.
The first part contains 5977 frames in total, recorded while the vehicle was switched off and connected to an external battery pack.
During this recording, the ambient temperature was approx. 25$^\circ$C (77$^\circ$F), which is similar to the temperature in the course of the second part acquisition, where 4645 frames were recorded.
Here, the vehicle was also switched off and in cold condition, but without having the additional battery pack connected.
The third part of the first data set consists of 6823 frames and was recorded during a 30 minutes long trip at a temperature of 32$^\circ$C (89.6$^\circ$F).
Afterwards, the vehicle was parked in a garage at approx. 23$^\circ$C (73.4$^\circ$F) for three hours with the intention to cool down the vehicular components.
For the recording of the fourth part, consisting of 4950 frames, the switched off vehicle was measured at first and subsequently driven for about 20 minutes at 36$^\circ$C (96.8$^\circ$F). 
In total, the first data set comprises 22,395 CAN frames and contains signals from a period of 5 hours and 15 minutes at a temperature range between 23$^\circ$C (73.4$^\circ$F) and 36$^\circ$C (96.8$^\circ$F).

The second data set contains a total of nine journeys over a period of five days, each lasting around 15 minutes on average.
The ambient temperatures were in the range of 10$^\circ$C (50$^\circ$F) on the first measuring day and just below 0$^\circ$C (32$^\circ$F) on the last.
Concerning the weather conditions during the recording period, the first day was for the most part rainy, on the second day it was still humid, while from the third day on the environment was dry.
These conditions facilitate our intention to use various electrical consumers, such as windscreen wipers, turn signals, lights, window regulators, heating, etc. during all journeys in order to record a potential impact on the voltage characteristics.
In addition, the measurements of the second data set include several vehicle starts, since the start-stop system was activated during the trips.
Altogether a total of 54,933 frames were recorded during these nine trips.
A summarized overview of both data sets which are used for the evaluation of this work is given in Table \ref{tab:data_sets}.

Between these two data sets lies a period of 19 weeks during which the vehicle was used by colleagues.
Further, a maintenance was carried out in this time, where the radiator fan and the battery of the Fiat were replaced.
Finally, it must also be noted that the measurement setup could not be 100\% identical for both data sets since it had to be removed from the car during this time.

\begin{figure*}
	\centering
	\includegraphics[width=\linewidth]{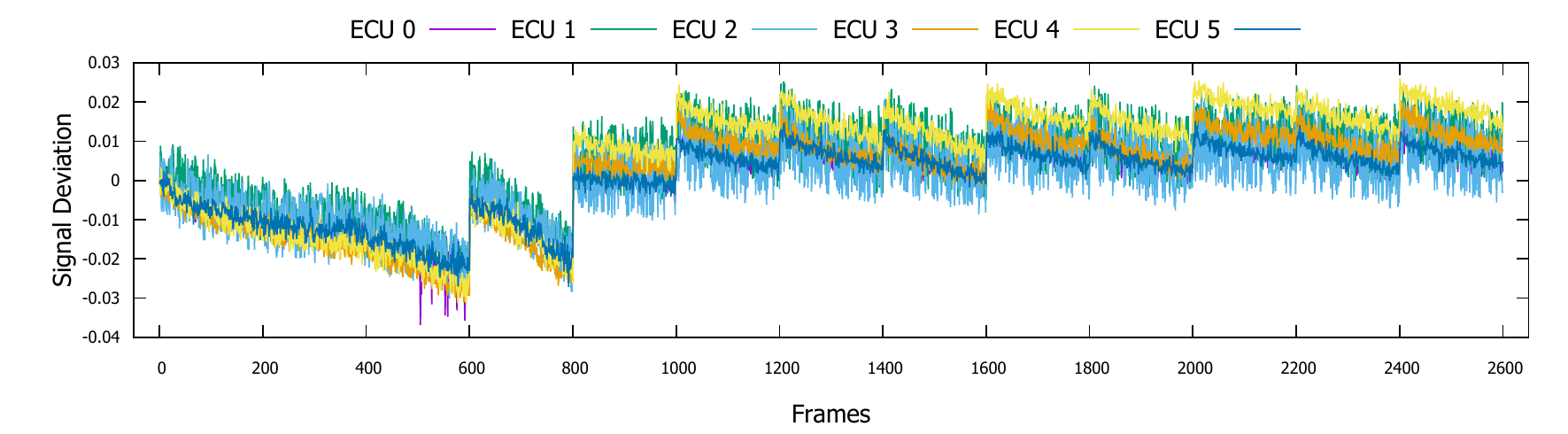}
	\caption{Signal deviations of the rising edges for all ECUs.}
	\label{fig_deviation_g10}
\end{figure*}

\subsection{Signal Analysis}
\label{subsec_signal_analysis}
\begin{figure}
		\centering
		
	\begin{subfigure}[c]{0.605\linewidth}
		\includegraphics[width=\linewidth]{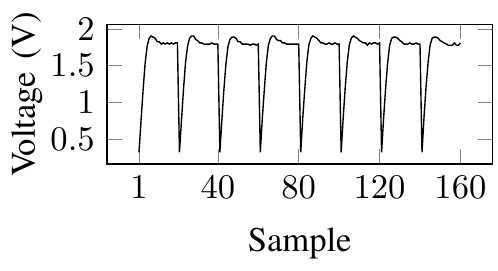}
		\caption{Group $G^{g,(e,m)}$ for $g=1$ with $K=8$ rising edges, i.e. symbols, for an arbitrary frame $m$ from ECU $e$.}
		\label{fig_rising_edge_group}
	\end{subfigure}
	\hfill
	\centering
	\begin{subfigure}[c]{0.38\linewidth}
		\includegraphics[width=\linewidth]{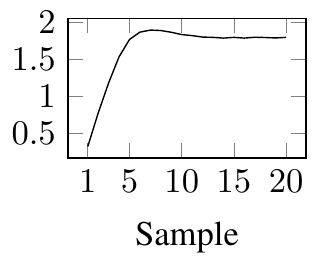}
		\caption{Average symbol of the opposite group $G^{g,(e,m)}$ for $g=1$.}
		\label{fig_rising_edge}
	\end{subfigure}
	\label{fig:prep_comparison}
	\caption{Preparation for the characteristics comparison}
\end{figure}

\paragraph{Signal Deviation Calculation}
Based on the introduced data set in the previous section, we now compare the voltage deviations between the recorded CAN messages of all individual sets, in order to get a feel for the change of voltage characteristics over a longer period of time.
To accomplish this intention, a method is required to compare individual messages which additionally operates independently of the transmitted data.
The foundation for this method is taken from the preprocessing step of the former presented Scission\,\cite{Kneib2018} approach.
In order to recall this step, Scission divides the sampled signal during the data transmission into three groups $g$ containing rising ($g=1$) and falling edges ($g=3$), as well as constant voltage levels during zero bit transmissions ($g=2$).
Note that in this section the classification and intrusion detection of Scission is not utilized but rather its approach to separate the voltage samples is taken up.
Figure\,\ref{fig_rising_edge_group} shows an exemplary group for $g=1$ containing $K=8$ rising edges.
Here, each rising edge, as well as a constant level or falling edge, is referred to as symbol.
Taking this into consideration, the $k$-th symbol of a frame $m$ coming from an ECU labeled with $e$ can be defined by 

\begin{equation}
S^{g,(e,m)}_k = (x_1,...,x_l) \label{eq:symbol}
\end{equation}

where, $x_i,\,i\in\{1,..,l\}$ are the individual voltage values of the symbol, $l$ their quantity per symbol and $g\in\{1, 2, 3\}$ the respective group.
Hence, with the help of Equation \ref{eq:symbol} we can also define a group $G^{g,(e,m)}$ of a frame $m$ coming from ECU $e$ by 

\begin{equation}
G^{g,(e,m)} = \bigcup_{k=0}^K S^{g,(e,m)}_k \label{eq:group}
\end{equation}

with $K$ being the number of symbols contained in the frame $m$ in respect to the group $g\in\{1, 2, 3\}$.
Here, it has to be noted that the data transmitted by the individual frames differs, leading to a varying number of bit transitions and thus to a different amount of symbols $K$ per group $g$ and frame $m$.
Therefore, the first step of comparing individual messages depicts the creation of an average symbol $\overline{S}^{g,(e,m)}$ for each group $G^{g,(e,m)}$ and frame $m$ in respect to $g\in\{1, 2, 3\}$, like shown in Figure\,\ref{fig_rising_edge}.
To achieve this, the samples are averaged over all $K$ symbols in $G^{g,(e,m)}$ calculated by

\begin{equation}
\overline{S}^{g,(e,m)}=(\frac{1}{K}\sum_{k=0}^{K}S^{g,(e,m)}_k[i]\mid\forall i\in\{1,...,l\}).
\label{eq:average_symbol}
\end{equation}


For the next step of the signal changes evaluation we use an adapted variant of the approach presented by Murvay and Groza\,\cite{Murvay2014}. 
First, a so-called template $\hat{S}^{g,e}$ is created for each of the three groups $g$ per \ac{ECU} $e$ by using the average symbols $\overline{S}^{g,(e,m)}$ of $M$ recorded messages originating from it.
Utilizing Equation \ref{eq:average_symbol} and a number $M$ of different frames, the template can be calculated by 

\begin{equation}
\hat{S}^{g,e}=(\frac{1}{M}\sum_{m=0}^{M}\overline{S}^{g,(e,m)}[i]\mid\forall i\in\{1,...,l\}).
\end{equation}
\label{eq:template}

It can be said that a template is the average of several previously mentioned average symbols.
The following frames of both data sets are then compared with the determined template.
This is achieved by calculating the average of the percentage difference per sample of the template and the average symbol of the message to be evaluated.
Mathematically, this signal deviation $SD^{g,(e,t)}$ of a data set frame $t$ originating from ECU $e$ in respect to the group $g$ can be expressed by 

\begin{equation}
SD^{g,(e,t)} = 1 - \sum_{i=0}^{l}\frac{\overline{S}^{g,(e,t)}[i]}{\hat{S}^{g,e}[i]}
\end{equation}

By calculating these deviations quotients over the course of the available data set, the concept drift captured over a longer period of time can now be analyzed.

\paragraph{Concept Drift Analysis}
\label{subsubsec_concept_drift}
The \ac{DSO} used to sample the CAN channel during data transmission has a sampling rate of 125\,MS/s, while Scission and its preprocessing step target a rate of 20\,MS/s.
Therefore, it was necessary to reduce the data points of all recorded signals by a factor of 3 by removing intermediate samples.
Afterwards, a template was created from $M=20$ messages per \ac{ECU} following the procedure described in the previous paragraph and used to calculate the voltage deviation of the remaining frames.
The general course of the signal deviation for every ECU of the Fiat\,500 is visualized in Figure\,\ref{fig_deviation_g10}. 
For the analysis we use the deviations of the rising edges, since these contain the most useful information for differentiation of the \acp{ECU}\,\cite{Kneib2018}.
Since the number of CAN frames per ECU and trip differs from each other, the figure shows the deviation of 200 frames per ECU and trip for a better illustration and comparison.

It is directly noticeable that the general course of the deviations between the \acp{ECU} is nearly identical.
This is also shown by the statistical evaluation of the signal deviations, presented in Table\,\ref{tab:data_stats}, showing an average correlation of 87\,\%.
There is no completely different course recognizable at any time, only the absolute values and the ranges of fluctuation vary.
This can be seen from the maximum difference between two succeeding frames, which are between 2.08$*10^{-2}$ and 3.47$*10^{-2}$.
But also the differences in the mean of the deviations e.g. for ECU\,4 with 1.49$*10^{-2}$ and ECU\,5 with 0.78$*10^{-2}$ indicate the fluctuation.
While stress on the battery will equally affect the signal characteristics of each \ac{ECU}, their different positions in the vehicle and thus on the CAN bus could have had a greater influence on these variations.
Since the deviations show an approximately similar course without noticeable differences, it can be assumed that the decisive electronic components of a CAN node have an approximately equal temperature gain.
Generally, it is recognizable that the deviations gradually increase with the duration of a journey, while they decrease abruptly after standstill phases.

\begin{figure*}
	\centering
	\includegraphics[width=\linewidth]{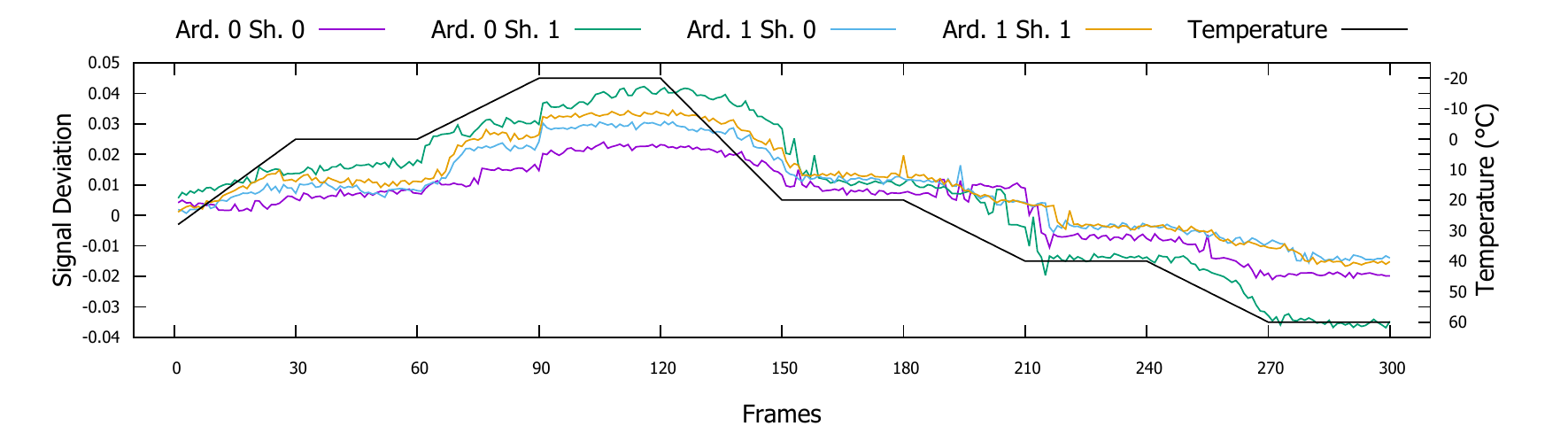}
	\caption{Dependency of signal deviations and temperature.}
	\label{fig_deviation_temp}
\end{figure*}

\begin{center}
	\begin{table}
		\caption{Statistical properties of the signal deviations.}
		\begin{tabular}{ccccccc}
			\hline \\[-1em]
			\multicolumn{7}{c}{Correlation}  \\ \\[-1em]
			& ECU\,0 & ECU\,1 & ECU\,2 & ECU\,3 & ECU\,4 & ECU\,5 \\
			\hline \\[-1em]
			ECU\,0 & 1 & .84 & .80 & .96 & .96 & .96 \\
			ECU\,1 &   & 1   & .71 & .85 & .85 & .85 \\
			ECU\,2 &   &     & 1   & .81 & .81 & .81 \\
			ECU\,3 &   &     &     & 1   & .98 & .81 \\
			ECU\,4 &   &     &     &     & 1   & .97 \\
			ECU\,5 &   &     &     &     &     & 1   \\
			& & & \\
			\hline \\[-1em]
			\multicolumn{7}{c}{Statistical Properties ($*10^{-2}$)} \\ \\[-1em]
			& ECU\,0 & ECU\,1 & ECU\,2 & ECU\,3 & ECU\,4 & ECU\,5 \\
			\hline \\[-1em]
			Mean      & 0.78 & 1.00 & 0.84 & 1.09 & 1.49 & 0.78 \\
			Deviation & 0.52 & 0.61 & 0.60 & 0.60 & 0.52 & 0.50 \\
			Maximum   & 3.69 & 2.54 & 3.03 & 3.10 & 3.12 & 2.71 \\
			Max. diff.  & 2.31 & 3.40 & 2.34 & 2.93 & 3.47 & 2.08 \\
		\end{tabular}
		\label{tab:data_stats}
	\end{table}
\end{center}

While the signals within the data sets, i.e. from frame 0 to 800 and from 800 to 2,600, have strong similarities in their course, there is a noticeable difference in  between.
As Figure\,\ref{fig_deviation_g10} indicates, there occurs the largest signal change of up to 3.69\,\%, mainly explainable by the repair measures of the vehicle between the two sets.
If in this case, the ambient temperature difference from summer to winter would depict the crucial factor, the signal levels of fully warmed up ECUs would have risen to a similar level at the end of each trip.
Comparing the measurement starts of the first and the last parts of the second data set, however, the ambient temperature cannot be completely neglected, as a very slight rise in the voltage level is also noticeable with a falling ambient temperature.
This can generally be explained by the increasing conductivity of electronic components at decreasing temperatures.
Basically, the analysis on occurring signal deviations makes apparent that repeating concepts are noticeable, while within the trips incremental concept drift due to temperature changes and in between the trips abrupt concept drift owing to aforementioned standstill phases and repair measures can be recognized.

\paragraph{Temperature Dependency}
\label{subsubsec_temp_depen}
We have performed a further experimental test to verify the assumption that the signal deviation is temperature-dependent.
Two Arduino Unos, each equipped with two CAN shields, were analyzed with the use of a Heraeus V\"otsch climatic chamber at a temperature range of -20$^\circ$C (-4$^\circ$F) up to 60$^\circ$C (140$^\circ$F).
The wiring of the \ac{CAN} bus, as well as the probes of the \ac{DSO} were also in the chamber during the measurements.
The Arduinos were powered by an external USB power supply.
It has to be made clear that the construction differs a lot from a real vehicle.
This applies not only to the missing bodies and the supply network, but also very strongly to the actual electronic components.
Such a simple test setup cannot represent a vehicle, but it can show a general tendency.

The analysis of the signal deviation, shown in Figure\,\ref{fig_deviation_temp}, was performed over a period of 3 1/2 hours, whereas 30 frames per temperature range are shown for better visualization.
The evaluation confirms the observation of the temperature dependence made during the analysis of the Fiat data.
It is clearly visible both in the illustration and in the correlations, which are shown in Table\,\ref{tab:data_stats_temp}.
As the temperature falls, the general voltage level of the signals increases and thus the difference deviates into the positive range. 
With rising temperatures, the course of signal differences behaves vice versa.
What is noticeable here, however, is that although there are similarities in the behavior of the shields equipped on the same Arduino, they do not necessarily have to be stable.
More precisely, while the deviation of the Arduino\,1 shields shows a very similar behavior and course, the Arduino\,0 shields have a greater difference.
The correlations between the \acp{ECU} also confirm this observation. 
While the correlations between the Arduino\,1 shields are 99.43\,\%, the Arduino\,0 shields show the same correlation as when the shields are compared between the two Arduinos.
From this observation it can be concluded that a construction of a general model for temperature-dependent change compensation is rather difficult, since each individual \ac{ECU} has a different behavior and would therefore require a separate measurement in advance.

\begin{center}
	\begin{table}
		\caption{Analysis of the temperature correlation.}
		\begin{tabular}{ccccc|c}
			\hline \\[-1em]
			\multicolumn{6}{c}{Correlation}  \\ \\[-1em]
				   & A\,0 S\,0 & A\,0 S\,1 & A\,1 S\,0 & A\,1 S\,1 & Temperature \\
			\hline \\[-1em]
			A\,0 S\,0 & 1     & .97    & .96   & .96   & -0.90\\
			A\,0 S\,1 &       & 1      & .97   & .97   & -0.95\\
			A\,1 S\,0 &       &        & 1     & .99   & -0.92\\
			A\,1 S\,1 &  	   &   		& 1     & 1     & -0.93\\
		\end{tabular}
		\label{tab:data_stats_temp}
	\end{table}
\end{center}

\subsection{Update Mechanism}
\label{sec:update_mechanism}
During the learning phase of a machine learning approach, a model is established based on the data of a fixed period in time.
After this establishment phase, it is very likely that the operating conditions change in the course of deployment due to a dynamic environment. 
As discovered in the previous section, the voltage characteristics are especially affected by external influences leading to the occurrence of both incremental or abrupt drift. 
With the purpose to preserve the performance of the system during these dynamically changing conditions, the model has to be adapted accordingly.

To enable a controlled adaption of the outdated model, the concept drift has to be recognized by the system in the first place.
During the initial model establishment, an average sender identification probability is determined which is used as reference for the operating phase of the system.
In the course of operation, the classification probabilities used for sender identification and intrusion detection are also utilized for the performance evaluation and are therefore continuously compared against the initially determined reference.
Should it be the case that the actual sender identification probabilities deteriorate by a certain degree, the necessity of a model update is indicated. 
By using an appropriate machine learning approach which is capable of establishing a model with incrementally available data, like it is the case with Logistic Regression, a model update can simply be performed with current training data. 

These update frames can either be continuously collected beforehand during the usual operation of the system in a bounded update set and immediately be used or can be gathered after the detection, which however delays the adaption procedure by a certain extent.
No matter when the frames are collected, it is of greater importance to include only trustworthy data into an update procedure. 
The reason for this is the fact that the operating environment of the system is rarely a trustworthy one, where an uncontrolled usage of transmitted frames for update purposes would enable an adversary to inject malicious frames into the adaption procedure~\cite{Huang2011}.
This adversarial sampling would allow the attacker to bypass the intrusion detection sooner or later.
Therefore, only frames which were classified with a high probability by the system are considered for an inclusion into the update set.

The aforementioned acquisition of update frames, however, is only applicable for incrementally changing voltage characteristics.
An abrupt change of the voltages, like seen in between the individual trips in Figure\,\ref{fig_deviation_g10}, can result in a to strong alteration of the signal, which can lead to the fact that no transmitted frame can be classified with a high probability.
This exacerbates the selection of trustworthy frames for the update set.
An option to bypass this issue depicts to explicitly secure the CAN frames with the goal of ensuring the authenticity.
Only in this way it can be assumed that an insertion of malicious frames into the update procedure does not take place.
With this intention, the temporary utilization of \acp{MAC} could be considered.
Although the application of \acp{MAC} has several drawbacks in the domain of CAN security, like bandwidth limitations and the requirement on higher computing capabilities, \acp{MAC} depict a viable option to provide trustworthy and labeled frames, which can be used to update the model and to handle abrupt voltage changes.
The difference in its usage to build a training set is that the cryptographic measurements are rarely and not continuously used and there is no need to transfer real-time critical data.
Furthermore, the available bandwidth is only slightly reduced by event-based messages, and not during the entire communication.
On a 65\,\% loaded bus at a baud rate of 500\,kb/s roughly 3000\,frames/s are transmitted. By e.g. 16 additional frames per ECU for updating the models, i.e. a total of 96 frames for the Fiat~500 considered here, the load of the bus would increase e.g. for one second by 2.08\,\%.
By the continous application of MACs for the protection of the communication, the usage of a MAC tag length of 24\,bits and 8\,bits for the freshness value would halve the bandwidth.
The resulting load of 130\,\%, since twice the amount of frames is necessary, would exceed the maximum, which is why the continuous use of MAC for the assumed network is not possible.
For this reason, the situational use compared to the continuous use of MACs represents only a minor effort.
Having updated the system with the most present and trustworthy data, its classification performance should be improved to the initial level until the next voltage deterioration is detected and the respective update procedure initiated.

\section{Evaluation}
\label{sec_evaluation}
In this section, we evaluate the different strategies in terms of performance and robustness.
First, we analyze whether a robust sender identification can be kept up over the entire training data without having to perform an update.
Initially the models are trained and then used to classify the remaining frames.
Afterwards the update technique is analyzed, as described in Section\,\ref{sec:update_mechanism}.
A model is trained initially, which is monitored during runtime and updated if necessary.
In order to visualize the performance of the strategies, the course of four different parameters is presented.
The \textit{Identification} indicates how many frames could be assigned to the correct sender. 
This provides direct information about the general performance of the sender identification.
The values \textit{True Positive} and \textit{True Negative} indicate how the system behaves when deciding between malicious and trustworthy frames.
For this purpose we have set the lower threshold to 0.2 and the upper threshold to 0.8.
A total of 5\,\% of the frames were treated as attacks, which was achieved by adjusting the identifiers while parsing the frames. 
The impersonated ECU has been continuously modified to ensure that each ECU is counterfeited by each ECU.
Finally, the \textit{Confidence} describes the general behavior of the model, which consists of all probabilities of the predictions.

\subsection{Initial Approach}
\label{sec_eval_initial}
\begin{figure}[ht]
	\centering
	\includegraphics[width=\linewidth]{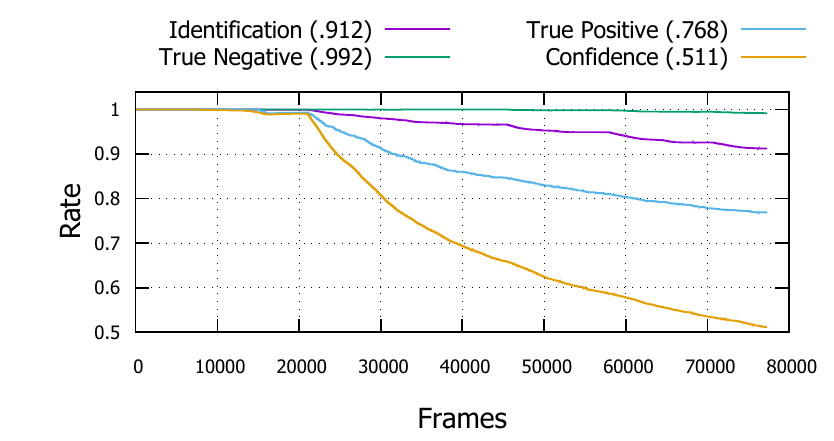}
	\caption{Performance at a training size of 200 frames.}
	\label{fig_eval_200}%
\end{figure}

To begin with, a static model is examined which is established with a small amount of training data.
For this, the model was initially trained with the first 200 frames per \ac{ECU}, as suggested in Scission\,\cite{Kneib2018}.
Subsequently, the remaining frames were processed by the system, i.e. the sender was determined and it was decided whether the frame was malicious or trustworthy.
The performance of this training approach is shown in Figure\,\ref{fig_eval_200}.

Figure\,\ref{fig_eval_200} shows that the average confidence level of the model starts to decrease slightly before frame 10,000. 
This point marks the transition from the second to the third part of the first data set, i.e. the shift from the switched-off to the switched-on vehicle.
From this point on, the performance of the system begins to decrease, especially sharply after frame 22,000.
The same applies to the number of true positives, which means that the chance of sending a counterfeit frame unnoticed increases.
This is due to the deterioration of the model, indicated by the confidence rate.
Due to the low probability values of the classifiers, it is less common for an attack to be marked as such, since the upper threshold for reducing false positives is rarely exceeded.
At the same time, however, it is also apparent that this greatly reduces the number of false positives.
Without the double threshold technique, the True Negative Rate would be equal to the Identification Rate, i.e. 91.23\,\%.
It can also be seen that the first false positives occur after frame 45,000.

\subsection{Increased Training Size}
\begin{figure}[ht]
	\centering
	\includegraphics[width=\linewidth]{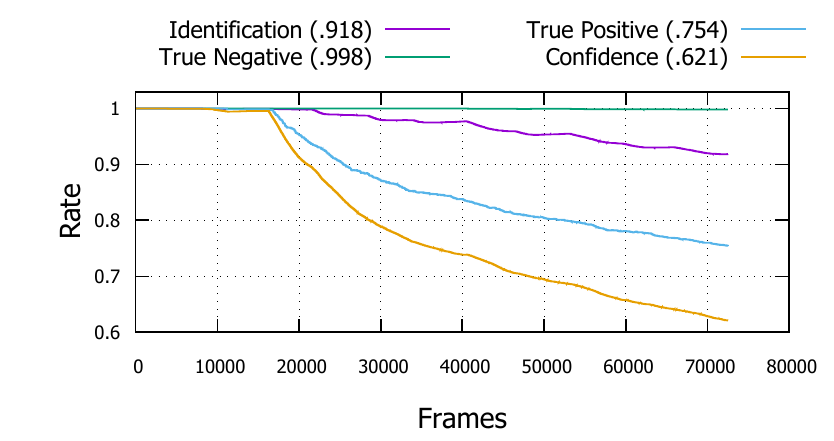}
	\caption{Performance at a training size of 1000 frames.}
	\label{fig_eval_1000}
\end{figure}

In this configuration it was examined whether a greater number of frames results in a noticeable improvement of performance and robustness.
Therefore, the model was trained with the first 1,000 frames per \ac{ECU}, instead of the first 200 frames, and then used for the evaluation of the remaining frames.
The behavior is illustrated in Figure\,\ref{fig_eval_1000}.

The results show that an increased amount of training leads only to a very slight improvement but not to a sufficiently good result. 
Although the false identifications are reduced by 7.3\,\% compared to the previous method, the performance is far behind the general identification rates\,\cite{Kneib2018} of 99.85\,\%.
Essentially, the decrease in performance is shifted by the additional training frames.
What is noticeable, however, is that the confidence of the model is higher.
In particular, the risk of confusion is about 22\,\% lower, since the probabilities of those \acp{ECU}, which are not the sender, are lower.
This also explains the increased True Negative Rate, as the probabilities of normal frames are less likely to be below the lower threshold. 
This reduces the number of frames marked as suspicious and at the same time reduces the chance of being misclassified as malicious.
Identification and True Positive Rate remain approximately the same.
The differences are primarily due to the low portion of frames of the first data set, since additionally 4,800 frames were used for training.
Therefore, we conclude that an increased number of training frames can enhance the performance and robustness slightly, but is not sufficient to deal with major changes.

\subsection{Update Mechanism}
\begin{figure}[ht]
	\centering
	\includegraphics[width=\linewidth]{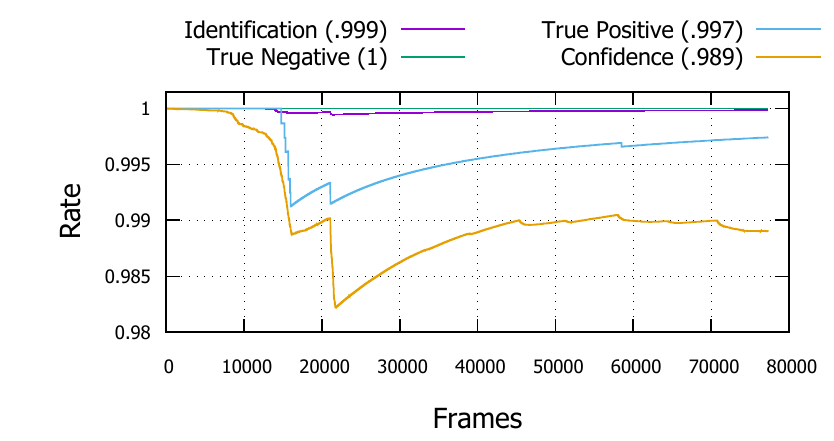}
	\caption{Performance using the update mechanism.}
	\label{fig:eval_update}
\end{figure}

Having compared the course of the performance for sender identification and intrusion detection for a static approach without an adaptation in the previous sections, we now include the update mechanism described in Section \ref{sec:update_mechanism} into the approach.
Here, the basis for the evaluation on the sender identification with the built in update mechanism is similar to the one without.
Concretely, a model is established on the basis of a fixed sized training set, in this case containing 200 frames per ECU. 
If a deterioration of the average probability is observed, the model is updated with the latest CAN frames.
In concrete terms, when the confidence for an ECU falls below the upper threshold of 0.8, an update is initiated.
For a single model update we have used 16 frames per \ac{ECU}.
This procedure is performed for the remaining test frames of the available data sets.
The results are shown in Figure\,\ref{fig:eval_update}.

The rates show the same slight performance decrease after frame 10,000.
At this point the chance to send a faked frame unnoticeably increases.
However, it is important to note here that the scale of the figure is significantly smaller than in the previous sections. 
Since the confidence of the individual classifiers is continuously monitored, the system recognizes its decrease and can thus react with targeted updates of individual classifiers.
In Figure\,\ref{fig:eval_update}, the first classifier is updated at frame 16,097, which results in an immediate increase in performance.
After changing the data set at frame 22,000 from summer to winter, another update of the classifiers is necessary. 
At this point, \acp{MAC} have to be used for the acquisition of reliable learning material, since after the four-month break and repair, a sufficiently high probability of classification could no longer be achieved. 
In concrete terms, this means that the confidence for the appropriate frames did not reach the upper threshold.
Without the use of \acp{MAC}, it cannot be ensured that the frames used to adapt the model are not manipulated.
Without this guarantee, poisoning attacks would be possible.
Alternatively, a targeted update of the system could be carried out in the workshop after the repair.
Nonetheless, after updating the model, it is visible that the overall performance of the system increases and that a high detection and identification rate can be achieved.
We could prevent all false positives and achieved a high classification probability of 99.98\,\%.
A large part of the undetected malicious frames are unnoticed as they were present exactly at the moment of the data set change, i.e. the time of a decreasing classification accuracy.
If these frames can be labelled e.g. after a longer standstill using \acp{MAC}, they can also be prevented and the True Positive Rate further increased.
Basically, it can also be seen here that an initial training to classify all frames of the large winter data set would be sufficient to determine all \acp{ECU} correctly without another update.
This is also confirmed by the results, if the experiment from Section\,\ref{sec_eval_initial} is only performed with the winter data set.

\subsection{Randomized Training Set}
\begin{figure}[ht]
	\centering
	\includegraphics[width=\linewidth]{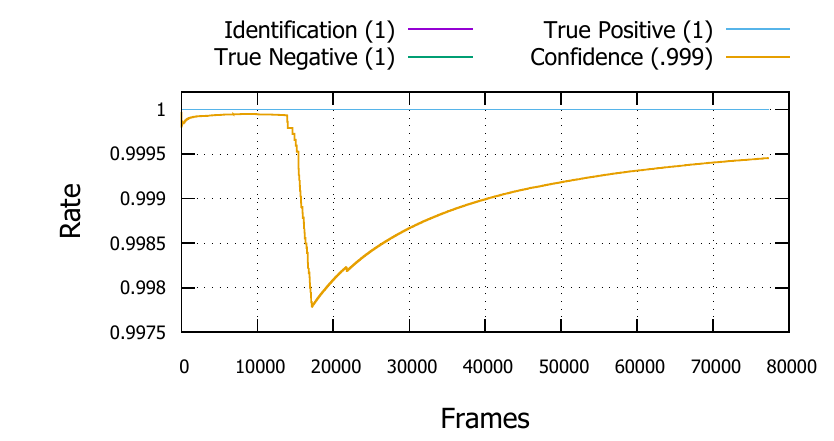}
	\caption{Performance using randomly choosen training frames.}
	\label{fig:eval_shuffle}
\end{figure}

It is common to use shuffled data for the evaluation of machine learning applications\,\cite{Rebala2019}, which is however difficult to implement for this technology.
Therefore, Scission uses the first frames per \ac{ECU} on the bus.
In order to understand the consequences of this restriction, we additionally investigate the behavior when 200 randomly selected frames are used per \ac{ECU} to train the model.
The remaining frames stay in the order they were measured on the bus.
The results are shown in Figure\,\ref{fig:eval_shuffle}.

To get rid of random noise, we repeated this test 20 times. 
Basically, the results were comparable in every run, the Confidence Rate falls slightly at frame 15,000.
This range contains the largest deviation of the signal as well as a few outliers, which corresponds to the range 500-600 in Figure\,\ref{fig_deviation_g10}. 
For the True Negative Rate there are no deviations, i.e. no false alarms occurred.
For the Identification and True Positive Rate there are minimal differences, as there are five runs of one false negative each.
This false estimation also occurs in the area with the largest deviations.
Overall, an average false negative probability of 0.00125\,\% was achieved.

\subsection{Parallel Measuring}
\begin{figure}
	\centering
	\includegraphics[width=\linewidth]{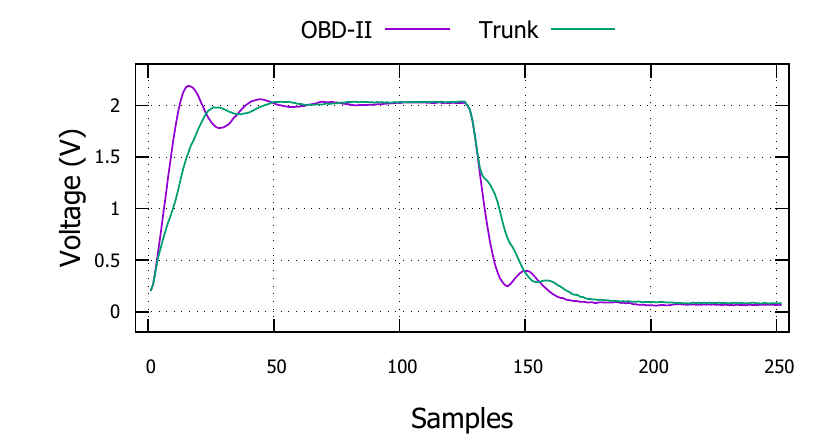}
	\caption{Exemplary parallel measuring difference.}
	\label{fig_double_meas}
\end{figure}

In Scission\,\cite{Kneib2018} it is mentioned that a mutual validation can be used to increase the reliability if the bus is observed from different measuring points.
In order to achieve a real advantage by monitoring the bus in parallel, the two models must have a certain difference.
Since the topology has an influence on the signal characteristics and the measuring point is also a part of this, its positioning also has an effect on the characteristics.
In order to investigate this fact we made another measurement trip.
We recorded 9,985 frames parallely at two points, the \ac{OBD}-II port and a direct access in the trunk of the vehicle.
By using several channels of the \acp{DSO} it was ensured that the identical frames were recorded.
It can be seen directly from the signals that the different measuring points contribute to the distinguishability of the characteristic.
The differences are noticeable in the shape of the edges.
In Figure\,\ref{fig_double_meas}, the differences are shown exemplarily for one frame.

\begin{figure}
	\centering
	\includegraphics[width=\linewidth]{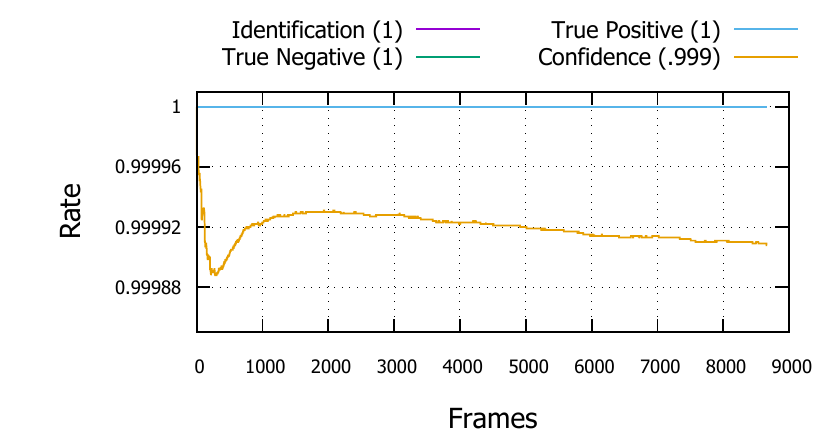}
	\caption{Performance measuring on OBD-II port.}
	\label{fig:eval_double_meas_a}
\end{figure}

\begin{figure}
	\centering
	\includegraphics[width=\linewidth]{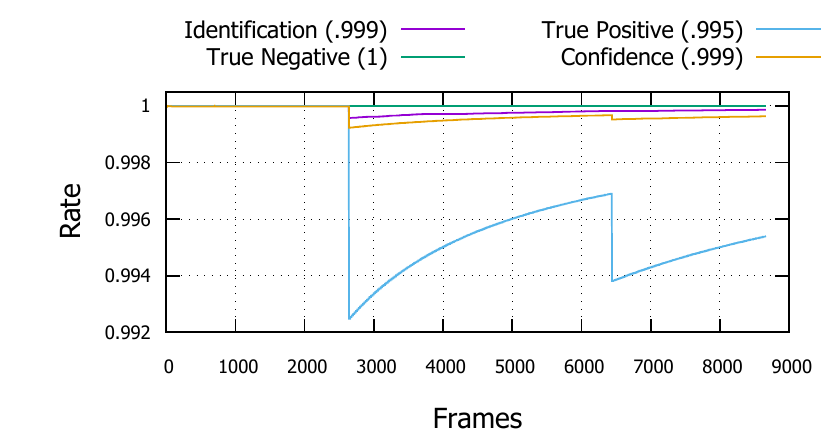}
	\caption{Performance measuring in the trunk.}
	\label{fig:eval_double_meas_b}
\end{figure}

A model with 200 frames was trained for both data sets to see if we can increase the robustness of the system. 
Subsequently, the remaining frames of the two instances were analyzed.
During the two evaluations the frames were exactly the same, only the measuring point was different.
The results are shown in Figure\,\ref{fig:eval_double_meas_a} and Figure\,\ref{fig:eval_double_meas_b}.
With identifcation rates of over 99.9\,\% and thus comparable with the results specified in Scission\,\cite{Kneib2018}, the performance is in principle comparably good for both measureing points.
Nevertheless, it is noticeable that the model, which was trained with the recordings from the trunk of the vehicle, did not recognize two manipulated frames.
However, the model which was trained with the measurements on the OBD-II port recognizes all manipulated frames.
As well as making the system more reliable to detect attacks, it also allows false positives to be double-checked to further reduce the chance of a false alarm.
Consequently, the statement that the reliability of sender identification systems can be increased by multiple verification at different measuring points is demonstrated and can thus be confirmed.

\subsection{Strategy Summary}
Fundamentally, we have shown that targeted strategies can greatly increase the performance and robustness of sender identification.
The tests confirmed the already recommended training size of 200 frames per \ac{ECU}.
An increased number does not achieve any significant improvements (see Figure\,\ref{fig_eval_200} and Figure\,\ref{fig_eval_1000}), but increases the effort for the training in terms of time, computing capacity and memory consumption.
Here, the utilization of an update mechanism provides advantages, but causes a further problem, since a secure training set cannot be set up with insufficient identification confidence.
The danger of poisoning attacks increases, hence the use of e.g. \acp{MAC} in order to create a labelled update set or training in a safe environment is necessary.
On the other hand, the performance is surprising when the model is trained with randomly selected frames (see Figure\,\ref{fig:eval_shuffle}). 
The results even exceed those of the update mechanism.
However, the implementation is difficult in practice, as the data per vehicle would have to be collected over several months.
But here, too, the secure labelling of the data must be guaranteed.
A combination is therefore considered to be appropriate.
Initially, the model can be trained with the first 200 frames per \ac{ECU}. 
For large changes, frames secured with \acp{MAC} are sent, which are used to update the model.
At the same time, a long-term training set can be set up in order to achieve the best possible general representation.
This set can then be used to replace the initial model.

\section{Related Work}
\label{sec_related_work}
\begin{figure*}[ht]
	\centering
	\includegraphics[width=\linewidth]{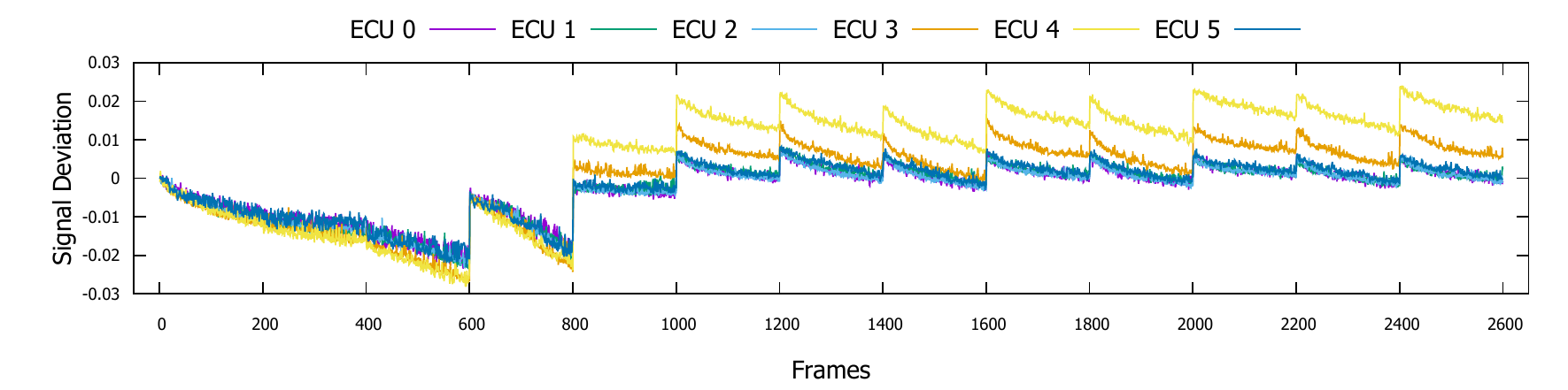}
	\caption{Signal deviations of the stable high signal for all ECUs.}
	\label{fig_deviation_g00}
\end{figure*}

During the operation of sender identification systems, different concept drifts occur as shown in Section\,\ref{subsec_signal_analysis}.
During a trip, mainly incremental drifts are observed, whereas in between there are abrupt or recurring drifts.
Basically, there are different strategies~\cite{Gama2014} for handling these.
However, due to the limited resources of the targeted hardware, these cannot be easily applied.
At the same time, the authenticity of learning material is of particular importance, as otherwise poisoning attacks are possible.

Even though Murvay and Groza\,\cite{Murvay2014} could not detect any concept drift in their prototype structure during a period of a few months, they note that these may occur and can be counteracted by continuously updating the model.
The work of Choi~\emph{et al.}\,\cite{Choi2016} contains no statements on a potential concept drift, whereas in the further development \cite{Choi2018} this is specifically analyzed. 
The researchers examined a real vehicle on two days with a temperature difference of 10$^\circ$C (50$^\circ$F).
The analyses support our observations that, among other things, temperature has an influence on the signals.
As a countermeasure, incremental learning is used here, which made it possible to maintain performance to a certain extent. 
However, there is a lack of research and information on how poisoning attacks can be prevented and how abrupt drifts can be handled in general.
In addition, incremental learning is very computation intensive, since each frame is used to update the model.

Viden\,\cite{Cho2017} does not use the entire signal for sender identification, but only the voltage level of high signals (e.g. sample 15 in Figure\,\ref{fig_rising_edge}).
In principle, this approach also continuously updates the model to compensate incremental changes.
This process is suspended when an attack is detected in order to prevent poisoning.
This requires an intact classification of the \acp{ECU} and therefore cannot be used for abrupt changes.
For these, Viden provides a common adaptation of the voltage profiles of all \acp{ECU}, since a similar voltage drop could be observed during a measurement.
Even if this worked in their presented evaluation, the results here (see Section\,\ref{subsubsec_concept_drift}) show that the signal changes are not equal across all \acp{ECU}.
This becomes even clearer if the deviations are not considered for the rising edge but for the stable high signal.
As it can be clearly seen in Figure\,\ref{fig_deviation_g00}, the deviations are different. 
While the deviations for \acp{ECU}\,0, 1, 2 and 5 are between frame 2,400 and 2,600 at a maximum of 0.6\,\%, the deviations for \ac{ECU}\,3 are between 0.6\,\% and 1.4\,\% and for \ac{ECU}\,4 between 1.5\,\% and 2.4\,\%.
A common adjustment of the voltage profiles can therefore lead to incorrect models, which means that the senders can no longer be reliably identified and hence poisoning of the model cannot be prevented.

\section{Conclusion}
\label{sec_conclusion}
Changes in analog signals need to be considered when used for sender identification. 
In our investigations, deviations of more than 3\,\% could be observed, characterized by incremental, abrupt and recurring changes of the signal characteristics, i.e.\ concept drifts.
In particular, a targeted experiment in a climatic chamber was able to demonstrate a high degree of temperature dependence.
By observing the performance of the sender identification methods, these changes can be detected and handled using an updating procedure.
Our updating procedure maintains general performance even with major changes in the operating conditions of the examined vehicle. 
Furthermore, with a long term update strategy, this work gives a recommendation for a continuously high system performance.
Our elaboration was also the first to examine the signal changes in greater detail and to use real data recorded over a longer period of time.
Together with the update procedure, this work closes the gap between sender identification procedures on the \ac{CAN} and its robust operation.
In summary, an identification rate and attack detection of over 99\,\%, as well as the avoidance of all false alarms could be achieved and shows for the first time that the operation of the technology is also possible with long-term signal changes.

\bibliographystyle{IEEEtran}
\bibliography{Robustness-arxiv}

\vfill

\end{document}